# Building a terminology network for search: the KoMoHe project

Philipp Mayr  Vivien Petras
GESIS Social Science Information Centre (GESIS-IZ), Bonn, Germany
philipp.mayr@gesis.org  vivien.petras@gesis.org

## Abstract

The paper reports about results on the GESIS-IZ project "Competence Center Modeling and Treatment of Semantic Heterogeneity" (KoMoHe). KoMoHe supervised a terminology mapping effort, in which 'cross-concordances' between major controlled vocabularies were organized, created and managed. In this paper we describe the establishment and implementation of cross-concordances for search in a digital library (DL).

**Keywords:** cross-concordances; terminology mapping; terminology service; subject searching

## 1. Project Background

Semantic integration seeks to connect different information systems through their subject metadata frameworks – insuring that distributed search over several information systems can still use the advanced subject access tools provided with the individual databases. Through the mapping of different subject terminologies, a 'semantic agreement' for the overall collection to be searched on is achieved. Terminology mapping – the mapping of words and phrases of one controlled vocabulary to the words and phrases of another – creates a semantic network between the information systems carrying the advantages of controlled subject metadata schemes into the distributed digital library world.

Terminology mappings could support distributed search in several ways. First and foremost, they should enable seamless search in databases with different subject metadata systems. Additionally, they can serve as tools for vocabulary expansion in general since they present a vocabulary network of equivalent, broader, narrower and related term relationships (see examples in TAB. 1). Thirdly, this vocabulary network of semantic mappings can also be used for query expansion and reformulation.

Starting point of the project was the multidisciplinary science portal vascoda[1] which merges structured, high-quality information collections from more than 40 providers in one search interface. A concept was needed that tackles the semantic heterogeneity between different controlled vocabularies (Hellweg et al., 2001, Krause, 2003).

In 2004, the German Federal Ministry for Education and Research funded a major terminology mapping initiative (KoMoHe project[2]) at the GESIS Social Science Information Centre in Bonn (GESIS-IZ), which found its conclusion at the end of 2007. One task of this terminology mapping initiative was to organize, create and manage 'cross-concordances' between major controlled vocabularies (thesauri, classification systems, subject heading lists) centered around the social sciences but quickly extending to other subject areas (see FIG. 1). The main objective of the project was to establish, implement and evaluate a terminology network for search in a typical DL environment.

In this paper, we describe the establishment and implementation of cross-concordances for search. A thorough information retrieval evaluation of several cross-concordances analyzing their effect on search was undertaken and is described in Mayr & Petras (2008 to appear).

---

[1] http://www.vascoda.de/
[2] http://www.gesis.org/en/research/information_technology/komohe.htm

## 2. Building a Cross-concordance Network

We define cross-concordances as intellectually (manually) created crosswalks that determine equivalence, hierarchy, and association relations between terms from two controlled vocabularies.

Typically, vocabularies will be related bilaterally, that is, a cross-concordance relating terms from vocabulary A to vocabulary B as well as a cross-concordance relating terms from vocabulary B to vocabulary A are established. Bilateral relations are not necessarily symmetrical. For example, the term 'Computer' in system A is mapped to the term 'Information System' in system B, but the same term 'Information System' in system B is mapped to another term 'Data base' in system A.

Cross-concordances are only one approach to treat semantic heterogeneity (compare Hellweg et al., 2001, Zeng & Chan, 2004).

Our approach allows the following 1:1 or 1:n relations:

- Equivalence (=) means identity, synonym, quasi-synonym
- Hierarchy (Broader terms <; narrower terms >)
- Association (^) for related terms
- An exception is the Null (0) relation, which means that a term can't be mapped to another term (see mapping number 4 in TAB. 1).

In addition, every relation must be tagged with a relevance rating (high, medium, and low). The relevance rating is a secondary but weak instrument to adjust the quality of the relations. They are not used in our current implementations. In our approach it takes approximately 4 minutes to establish one mapping between two concepts. Table 1 presents typical unidirectional cross-concordances between two vocabularies A and B.

| No | Vocabulary A | Relation | Vocabulary B | Description |
|----|--------------|----------|--------------|-------------|
| 1 | hacker | = | hacking | Equivalence relationship |
| 2 | hacker | ^+ | computers + crime | 2 association relations (^) to term combinations (+) |
| 3 | hacker | ^+ | internet + security | |
| 4 | isdn device | 0 | | Concept can't be mapped, term is too specific. |
| 5 | isdn | < | telecommunications | Narrower term relationship |
| 6 | documentation system | > | abstracting services | Broader term relationship |

TAB. 1. Cross-concordance examples (unidirectional).

The mappings in the KoMoHe project involve all or major parts of the vocabularies. Vocabularies were analyzed in terms of topical and syntactical overlap before the mapping started. Term lists are precompiled and ready to map when they come to people who are mapping. Collaborative work on one mapping is possible, but more complicated to organize. All mappings are created by researchers or terminology experts. Essential for a successful mapping is an understanding of the meaning and semantics of the terms and the internal relations (structure) of the concerned vocabularies[3]. This includes syntactic checks of word stems but also semantic knowledge to look up synonyms and other related terms. See in this context Lauser et al. (2008, to be published) for an insight concerning intellectual and automatic mapping methodologies.

The mapping process is based on a set of practical rules and guidelines (see also Patel et al., 2005). During the mapping of the terms, all intra-thesaurus relations (including scope notes) are consulted. Recall and precision of the established relations have to be checked in the associated databases. This is especially important for combinations of terms (1:n relations). One-to-one (1:1) term relations are preferred. Word groups and relevance adjustments have to be made consistently.

---

[3] Some of the same problems occur in the development of multilingual thesauri, which are detailed in IFLA (2005) and the ISO 5964 (1985) standard.

In the end, the semantics of the mappings are reviewed by experts and samples are empirically tested for document recall and precision. Expert reviews focus especially on semantic correctness, consistency and relevance of equivalence relations which are our most important relationship type. Sampled mappings are cross-checked and assessed via queries against the controlled term field of the associated database.

More mapping examples can be found in Mayr & Walter (2008).

To date, 25 controlled vocabularies from 11 disciplines and 3 languages (German, English and Russian) have been connected with vocabulary sizes ranging from 1,000 – 17,000 terms per vocabulary (see the project website for more details). More than 513,000 relations were generated in 64 crosswalks. Figure 1 depicts the established network of cross-concordances by discipline.

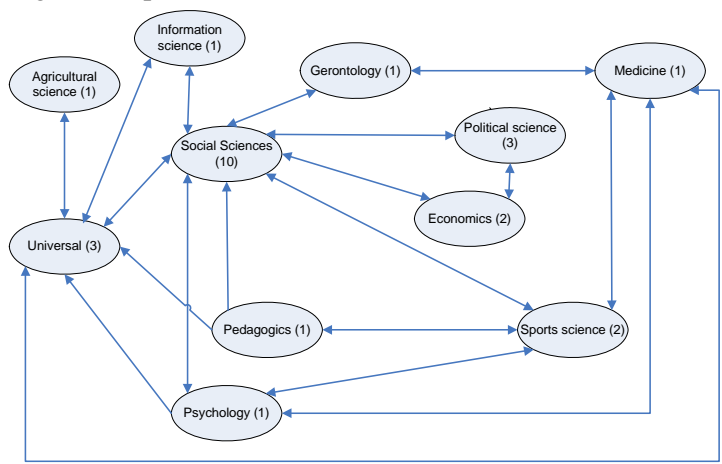

FIG. 1. Network of terminology mappings in the KoMoHe project. The numbers in brackets contain the number of mapped controlled vocabularies in a discipline.

## 3. Implementing Cross-concordances for Search

A relational database was created to store the cross-concordances for later use. It was found that the relational structure is able to capture the number of different controlled vocabularies, terms, term combinations, and relationships appropriately. The vocabularies and terms are represented in list form, independent from each other and without attention to the syndetic structure of the involved vocabularies. Orthography and capitalization of controlled vocabulary terms were normalized. Term combinations (i.e. computers + crime as related combination for the term hacker) were also stored as separate concepts.

To search and retrieve terminology data from the database, a web service (called heterogeneity service, see Mayr & Walter, 2008) was built to support cross-concordance searches for individual start terms, mapped terms, start and destination vocabularies as well as different types of relations.

Many cross-concordances are already utilized for search in the German Social Science Information Portal sowiport[4], which offers bibliographical and other information resources (incl. 15 databases with 10 different vocabularies and about 2.5 million bibliographical references). The application, which uses the equivalence relations[5], looks up search terms in the controlled vocabulary term list and then automatically adds all equivalent terms from all available vocabularies to the query. If the controlled vocabularies are in different languages, the heterogeneity service also provides a translation from the original term to the preferred controlled term in the other language. If the original query contains a Boolean command, it remains intact after the query expansion (i.e. each query word gets expanded separately). In the results list, a small icon symbolizes the transformation for the user (see FIG. 2).

---

[4] http://www.sowiport.de/

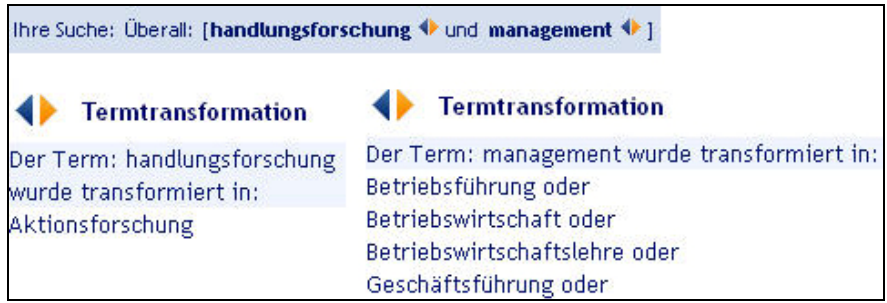

FIG. 2. Term mapping for search in sowiport. All terms are added to the query with a Boolean OR.

Because of performance issues, the cross-concordance query expansion doesn't distinguish between different databases and their preferred controlled vocabulary terms given a concept, but adds all equivalent terms to the query. In principle, this use of the terminology network expands a query with synonyms or quasi-synonyms of the original query terms. By adding terms to the query, recall should increase, that is, more relevant documents will be found. It is unclear, however, whether the indiscriminate expansion of the original query without regard for the terms' appropriateness for a given database can actually decrease the precision of the search. If the created equivalence mappings only denote correct synonyms, then the adding of true synonyms should have no such effect. However, homonymic terms as well as slight variations in the meaning of a concept can have a detrimental impact on the quality and precision of the query. In an ideal case, the searcher could be represented with a selection of terms garnered from the cross-concordances and then select an appropriate formulation. As most users prefer simple search interfaces with quick results (Jansen & Pooch, 2000; Bandos & Resnick, 2004), an interactive search process or even an appropriate visualization of the cross-concordance work is difficult to accomplish.

Another major issue for a growing terminology network is the scale and overlap of cross-concordances. The more vocabularies are mapped to each other, the more terms occur multiple times in variant mappings[6], which makes automatic query expansion more imprecise. On the other hand, the more vocabularies are added in such a network, the more inferences can be drawn for additional mappings. Indirect mappings via a pivot vocabulary could help in connecting vocabularies that haven't been mapped to each other. A sufficiently large network could assist in reducing the mapping errors introduced by statistical or indirect mappings.

## 4. Leveraging a Terminology Network – Outlook

This project is the largest terminology mapping effort in Germany. The number and variety of controlled vocabularies targeted provide an optimal basis for further research opportunities. To our knowledge, terminology mapping efforts and the resulting terminology networks have rarely been evaluated with stringent qualitative and quantitative measures.

---

[5] The other relations, which can lead to imprecise query formulations because they are broader, narrower or related to the original term, could be leveraged in an interactive search, when the searcher can guide and direct the selection of search term.

[6] For example: term A from vocabulary 1 also occurs in vocabulary 2. A variant mapping exists when term A from vocabulary 1 is mapped to term B in vocabulary 3, but term A from vocabulary 2 is mapped to term C in vocabulary 3. This might be the correct mapping because the concepts in the different vocabularies are differently connotated but most of the time this will introduce noise to the network.

The current cross-concordances will be further analyzed and leveraged for distributed search not only in the sowiport portal but also in the German interdisciplinary science portal vascoda. The terminology mapping data is made available for research purposes. Some mappings are already in use for the domain-specific track at the CLEF (Cross-Language Evaluation Forum) retrieval conference (Petras, Baerisch & Stempfhuber, 2007).

We also plan on leveraging the mappings for vocabulary help in the initial query formulation process as well as for the ranking of retrieval results (Mayr, Mutschke & Petras, 2008).

Aside from its application in a distributed search scenario, the semantic web community might be able to find new and interesting usages for terminology data like this one. The SKOS standard (Simple Knowledge Organization System)[7] contains a section on mapping vocabularies in its draft version. Once the standard gets stabilized, we plan on transferring the cross-concordance data to the SKOS format. If more vocabularies and mappings become available in SKOS, then further research into connecting previously unmapped terminology networks with each other should be possible.

## Acknowledgements

The project was funded by BMBF, grant no. 01C5953. We wish to thank all our project partners for their collaboration. We appreciate the help of our colleagues Anne-Kathrin Walter, who implemented the database and the heterogeneity service prototype, and Stefan Bärisch, who integrated the cross-concordances in the current sowiport search.

---

[7] http://www.w3.org/2004/02/skos/